\documentclass[aps,prb,twocolumn,superscriptaddress,showpacs,showkeys]{revtex4-1}

\usepackage{amssymb,amsmath,amsfonts}
\usepackage[mathcal]{eucal}
\usepackage{subfigure}
\usepackage{bm,mathrsfs}
\usepackage{hyperref}

\usepackage{tikz}

\input{epsf}

\newcommand{\ds}{\displaystyle}

\newcommand{\fp}[1]{{\ds\frac{\partial}{\partial #1}}}
\newcommand{\fpt}[2]{{\ds\frac{\partial #1}{\partial #2}}}

\newcommand{\Sp}{\mathop{\rm Sp}\nolimits}
\newcommand{\Tr}{\mathop{\rm Tr}\nolimits}
\newcommand{\tr}{\mathop{\rm tr}\nolimits}
\renewcommand{\Im}{\mathop{\rm Im}\nolimits}
\renewcommand{\Re}{\mathop{\rm Re}\nolimits}
\newcommand{\aver}[1]{\left\langle{#1}\right\rangle}
\newcommand{\cmatrix}[3]{\left\langle{#1}\left|{#2}\right|{#3}\right\rangle}

\newcommand{\slashnabla}{\rotatebox[origin=c]{180}{$\varDelta$}}

\newcommand{\cE}{\mbox{$\cal E$}}
\newcommand{\cG}{\mbox{$\cal G$}}
\newcommand{\cH}{\mbox{$\cal H$}}
\newcommand{\cN}{\mbox{$\cal N$}}

\newcommand{\tM}{\mbox{$\tilde M$}}

\newcommand{\sG}{\mbox{$\sf G$}}

\newcommand{\vsigma}{\mbox{\boldmath $\sigma$}}
\newcommand{\vpi}{\mbox{\boldmath $\pi$}}
\newcommand{\vrho}{\mbox{\boldmath $\varrho$}}
\newcommand{\vpiscr}{\mbox{\boldmath\scriptsize $\pi$}}
\newcommand{\vrhoscr}{\mbox{\boldmath\scriptsize $\varrho$}}

\begin{document}

\title{Topological robustness of quantization of the anomalous Hall conductance
\\of a two-dimensional disordered Chern insulator}

\author{S.\,G. Novokshonov}
\email[]{nov@imp.uran.ru}

\affiliation{M.\,N.~Mikheev Institute of Metal Physics of Ural Branch of Russian
Academy of Sciences, 620108, 18, S. Kovalevskaya st., Ekaterinburg, Russia}
\affiliation{Institute of Physics and Technology, Ural Federal University,
620002, 19, Mira st., Ekaterinburg, Russia}

\date{\today}

\begin{abstract}
The robustness of the anomalous Hall conductance, $\sigma_{yx}^{}=
\sigma_{H}^{}$, quantization in the model of a~two-dimensional disordered gas
of massive Dirac electrons subjected to an external orthogonal magnetic field
is investigated in the framework of Kubo--St\v{r}eda formalism. Using the 
momentum representation for the averaged one-electron Green functions in 
a~magnetic field, an explicit analytical expression for the St\v{r}eda term of
$\sigma_{H}^{}$ is obtained. It is shown that this term is proportional to the 
topological Chern number, ${\rm Ch}=\pm 1/2$, if the Fermi level is in the
energy gap. In this case, the total $\sigma_{H}^{}$ takes the half-integer
quantized value, $\sigma_{H}^{}=\pm e_{}^{2}/4\pi\hbar$,that does not depend
on either the magnitude of the disorder or the strength of the external
magnetic field. As an example, we calculated the densities of states and
intrinsic anomalous Hall conductance, $\sigma_{H}^{\rm int}$, of
a~two-dimensional disordered gas of massive Dirac electrons subjected to an
external magnetic field in the self-consistent Born approximation. A numerical 
analysis of the field and energy dependencies of these expressions is carried
out for various values of the parameters of the model under consideration. In 
particular, the results of this analysis show that the St\v{r}eda term of
$\sigma_{H}^{}$ is susceptible to disorder in a~sufficiently wide vicinity of
the point of transition to the quantization regime. 
\end{abstract}

\maketitle

\section{\label{sec:intro} Introduction}

In the last years, a widespread interest was attracted to investigation of the
quantum anomalous Hall effect (quantum AHE or QAHE) in the topological
non-trivial materials \cite{weng_etal_2015,wang_etal_2015,liu_etal_2016}. The 
possibility of quantizing Hall conductance without Landau levels was first 
demonstrated by Haldane \cite{haldane_1988} based on a~tight-binding model with
a~zero net magnetic flux through the unit cell of the honeycomb lattice. 
Subsequently, several authors predicted the quantization of anomalous Hall
conductance in two-dimensional (2D) magnets with spin-orbit interaction \cite
{ono_nag_2002,ono_nag_2003,cul_etal_2003,qi_etal_2006}. This effect was first 
discovered experimentally by Cui-Zu Chang with co-workers \cite{chang_etal_2013} 
in thin films of the $({\rm Bi}_{x}^{}{\rm Sb}_{1-x}^{})_{2}^{}{\rm Te}_{3}^{}$
topological insulator doped by Cr. An overview of recent experimental and
theoretical studies of QAHE can be found in review by Culcer et al.\cite
{cul_etal_2020}

Assuming $T=0$, the expression for the anomalous Hall conductance $\sigma_{H}^{}
=\sigma_{yx}^{}$ of a~clean electron system with broken time reversal symmetry 
can be written as \cite{ono_nag_2002,jung_etal_2002}
\begin{equation}
\label{eq:sigma_berry}
\sigma_{H}^{}=\sigma_{0}^{}\sum_{n}\!
\int_{{\cal E}_{n}^{}(\bm p)<{\cal E}_{F}^{}}\!\Omega_{n,z}^{}(\bm p)\,\frac
{d\bm p}{2\pi}.
\end{equation}
Here, $\sigma_{0}^{}=e_{}^{2}/2\pi\hbar$ is the conductance quantum,
$\Omega_{n,z}^{}(\bm p)=i\langle n\bm p|(\stackrel{\leftarrow}
{\slashnabla_{\!\!\bm p}^{}}\times\stackrel{\rightarrow}
{\slashnabla_{\!\!\bm p}^{}})_{z}^{}|n\bm p\rangle$ is the $z$-component of
Berry's curvature vector\cite{berry_1985} of the electron states $|n\bm p
\rangle$ manifold in $n$th energy band. The arrows above the nabla operators 
indicate the directions of their actions. 

If the Fermi level $\cE_{F}^{}$ is in an energy gap, then the integration in
the each non-zero summand in (\ref{eq:sigma_berry}) is performed over the
closed manifold without boundary. As a~result, these integrals become equal to
integers Chern numbers $\rm Ch$ (or $\pm 1/2$ for the model of 2D massive Dirac 
electrons), and the Hall conductance takes on a~quantized value that is 
a~multiple of $\sigma_{0}^{}$. In this regime, $\sigma_{H}^{}=\sigma_{0}^{}{\rm
Ch}$ is non-zero entirely due to the so-called intrinsic mechanism of AHE 
proposed by Karplus and Luttinger\cite{karp_lutt_1954}. This mechanism is
caused by non-trivial topology of the electron states in ideal
crystal\cite{cul_etal_2003,jung_etal_2002}. Therefore, the quantization of
$\sigma_{H}^{}$ must be protected against perturbations introduced by both
internal disorder and external fields, for example, magnetic field. The 
numerical finite-size scaling analysis of the anomalous Hall conductance of the 
Haldane model\cite{haldane_1988} with Anderson disorder confirms this
conclusion\cite{ono_nag_2003}. Namely, the $\sigma_{H}^{}$ tends to the 
quantized value $e_{}^{2}/2\pi\hbar$ in the thermodynamic limit if the Fermi 
level is in the mobility gap.

The fundamental problem of the influence of an external magnetic field on the
QAHE was investigated by B\"{o}ttcher et
al.\cite{boet_etal_2019,boet_etal_2020}.
Using the effective field theory, the authors of these works showed that
quantization of the anomalous Hall conductance  $\sigma_{H}^{\rm a}$ survives
in an external orthogonal magnetic field and can be distinguished from the
quantization of normal Hall conductance due to its parity $\sigma_{H}^{\rm a}
(B)=\sigma_{H}^{\rm a}(-B)$.

In both these cases, the quantized anomalous Hall conductance is proportional
to the Chern number. But the equation (\ref{eq:sigma_berry}) holds for a~free 
system in the clean limit when the momentum is a~well-defined quantum number. 
However, there is an alternative expression for the Chern number in terms of 
one-electron Green's functions
(GF)\cite{thoul_etal_1982,q_niu_etal_1985,volovik_1988,volovik_2003}, 
which can be generalized to the case of disorder, electron-electron interaction, 
and external magnetic field\cite{ishi_mat_1987,matsu_1987,zubkov_2016}. The
most general expression for the Chern number in terms of one-electron GF was 
obtained by Zubkov et al.\cite{zubkov_2016,zub_wu_2020,zhang_zub_2019} in the 
Wigner representation, which is convenient for studying spatially nonuniform 
systems.

Thus, the understanding of the physical nature of the quantization of the
anomalous Hall effect has reached a~fairly high level by now. Nevertheless,
there are a~number of problems in this area that need to be addressed. In
particular, the question of the influence of disorder and external magnetic
field on the quantization of the anomalous Hall conductance and its behavior in
vicinity of the plateau has not been sufficiently studied. The present work
is devoted to solving this problem. 

The main text of the article is organized as follows. Sec.~II describes a~model
of massive Dirac electrons for a two-dimensional disordered Chern insulator,
and also presents the Kubo-St\v{r}eda formulas for the Hall conductance of 
non-interacting disordered system. In Sec.~III, the calculation of the
self-energies and densities of states (DOS) of electrons necessary for the 
subsequent analysis is performed in the self-consistent Born approximation 
(SCBA). In Sec.~IV, we derive the relation between the St\v{r}eda term of the 
Hall conductance and the topological Chern number, which is valid in the 
presence of disorder and an external magnetic field. In Sec.~V, using the 
results obtained above, we calculate in SCBA the intrinsic anomalous Hall
conductance of a~two-dimensional disordered gas of massive Dirac fermions
subject to an orthogonal external magnetic field. Finally, we discuss our
results in Sec.~VI.

\section{\label{sec:model} Model}

Let us consider a two-dimensional (2D, $||\rm OXY$) disordered Chern insulator
subject to an external orthogonal ($||\rm OZ$) magnetic field $\bm B=
\slashnabla\times\bm A$. The one-particle Hamiltonian of the minimal model of 
the system under consideration has the following form\cite{bernevig_2013}
\begin{equation}
\label{eq:chern_ham}
\cH+U=\bm d_{\vpiscr}^{}\cdot\vsigma+U=v(\vpi\cdot\vsigma)+M\sigma_{z}^{}
+U(\bm r),
\end{equation}
Here, the mechanical momentum  $\vpi=\bm p-e\bm A/c$ of an electron is treated
as a~two-dimensional vector in the plane of the considered system, $\vsigma=
(\sigma_{x}^{},\sigma_{y}^{},\sigma_{z}^{})$ is the 3D-vector formed by the
Pauli (pseudo-) spin matrices, $v$ is a~model parameter, which has the meaning
of some characteristic velocity, and $M$ is proportional to Dirac mass $|M|
/v_{}^{2}$. The inclusion of the term $M\sigma_{z}^{}$ leads to violation of
the time reversal symmetry of the Hamiltonian $\cH$ (\ref{eq:chern_ham}) and
also opens the gap $\cE_{g}^{}=2|M|$ in its spectrum. Finally, $U(\bm r)$ is
a~white-noise Gaussian random potential field. 

\begin{figure}[t!]
\vspace*{-1.0cm}
\hspace*{-0.4cm}\includegraphics[scale=0.23]{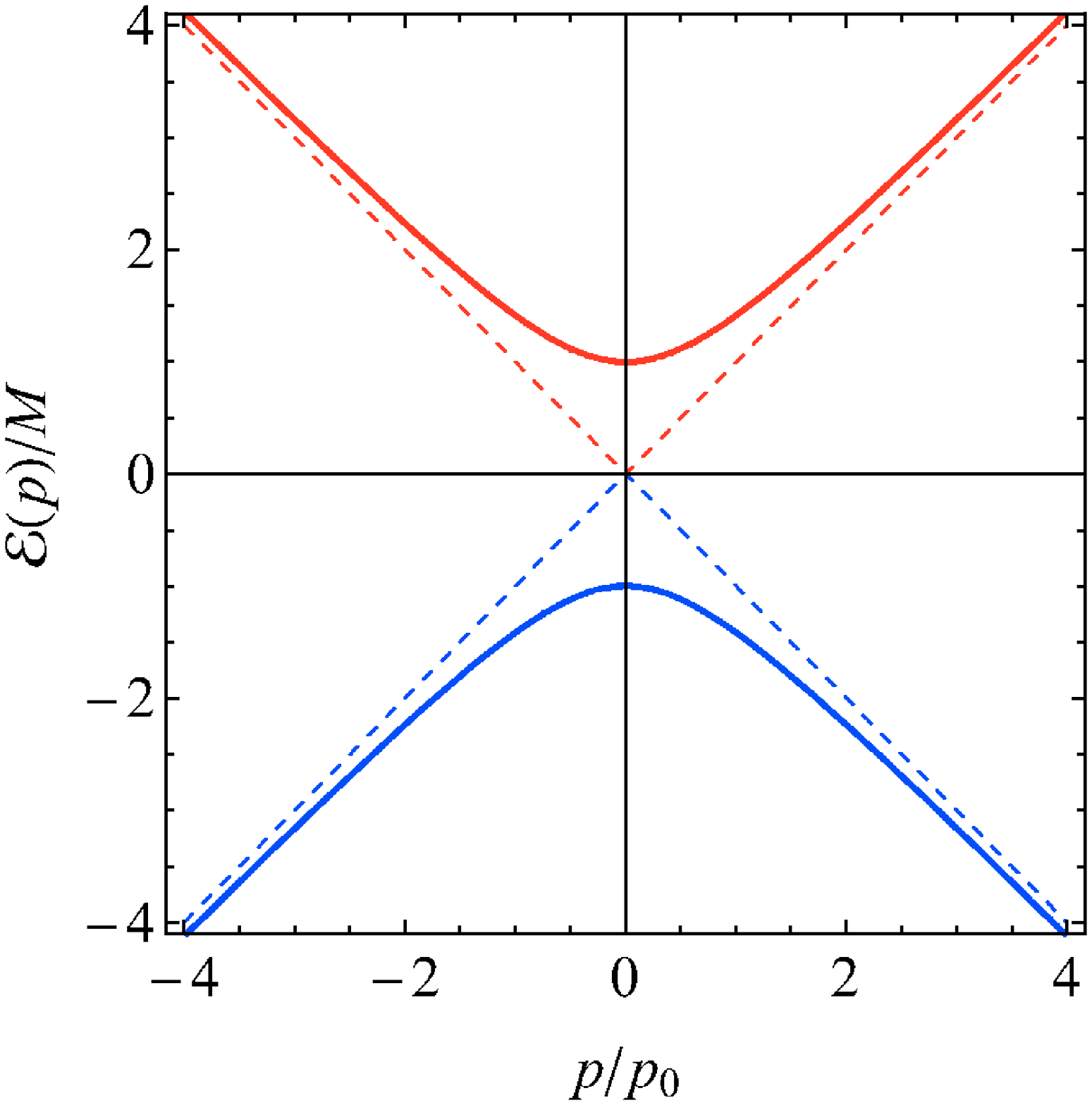}\hspace*{-0.7cm}
\includegraphics[scale=0.23]{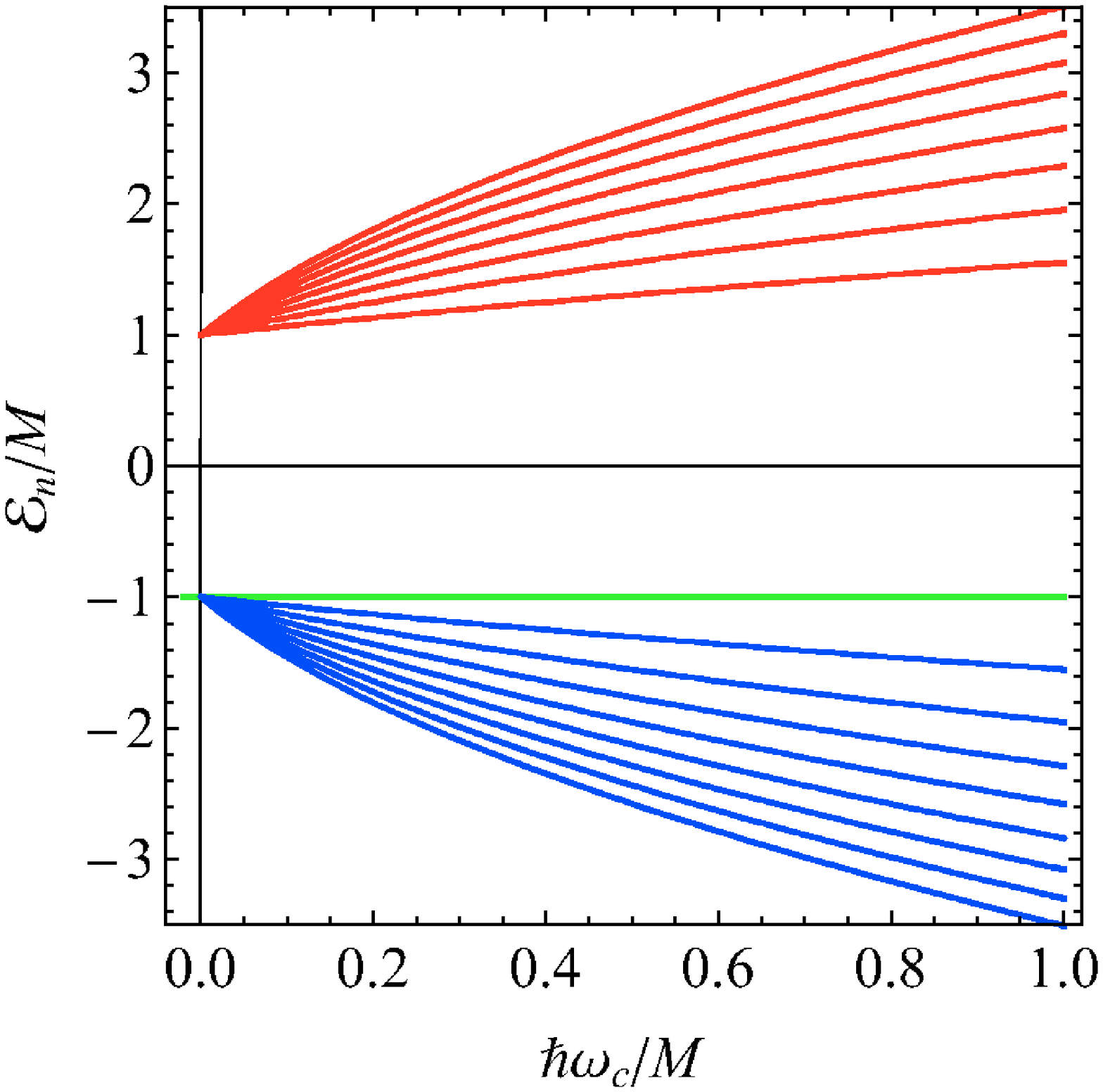}%
\vspace*{-1.2cm}
\caption{\label{fig:energy_spectra} (color online) \underline{Left panel}: The
gapped ($\cE_{g}^{}=2|M|$) one-particle dispersion $\cE_{s}^{}(\bm p)$ 
(\ref{eq:eigen_energy}). Dashed lines represent the gapless spectrum of
a~massless Dirac electron, $p_{0}^{}=|M|/v$.\\
\underline{Right panel}: The gapped ($\cE_{g}^{}=|M|+\sqrt{M_{}^{2}+2mv_{}^{2}
\hbar\omega_{c}^{}}$) series of the Landau levels (\ref{eq:landau_levels}). The
horizontal line depicts the anomalous Landau level $\cE_{-1,0}^{}=-M$ ($M>0$)
in the lower subband of the electron spectrum.}
\end{figure}

In absence of an external magnetic field, the electron dispersion in the
clean model (\ref{eq:chern_ham}) is the energy spectrum of two-dimensional
massive Dirac electrons
\begin{equation}
\label{eq:eigen_energy}
\cE_{s}^{}(\bm p)=s\cE(\bm p)=s\sqrt{M_{}^{2}+2mv_{}^{2}\cE_{\bm p}^{}},
\end{equation}
where $\cE_{\bm p}^{}=\bm p_{}^{2}/2m$ is the kinetic energy of a~free
non-relativistic electron. The quantum number $s=\pm 1$ determines the helicity
of the electron eigenstate with energy (\ref{eq:eigen_energy}), that is, the
projection of its (pseudo-) spin onto the direction of the vector
$\bm d_{\bm p}^{}$. In an orthogonal external magnetic field ($\bm B||\rm OZ$),
the spectrum (\ref{eq:eigen_energy}) is transformed into two series of Landau
levels
\begin{equation}
\label{eq:landau_levels}
\cE_{s,n}^{}=s\cE_{n}^{}=s\sqrt{M_{}^{2}+2mv_{}^{2}\hbar\omega_{c}^{}n}\,,
\quad~\begin{array}{l} s=+1,~~~n\geqslant 1,\\ s=-1,~~~n\geqslant 0,\end{array}
\end{equation}
where $\omega_{c}^{}=|e|B/mc$ is the cyclotron frequency of a free electron,
$n$ is the integer number of the Landau level $\cE_{n}^{}$.  The typical 
one-particle dispersion (\ref{eq:eigen_energy}) and two fans of the Landau
levels (\ref{eq:landau_levels}) are shown in Fig.~\ref{fig:energy_spectra}.

According to the Kubo--St\v{r}eda formalism \cite{streda}, the expression for
the Hall component of the conductivity tensor of the system under consideration
can be written as
\begin{align}
\label{eq:kubo_total}
\sigma_{H}^{}=&\,\sigma_{H}^{\rm I}+\sigma_{H}^{\rm II}=-\frac{\hbar e_{}^{2}}
{2\pi}\!\int\!\fpt{f}{\cE}\Tr\big\langle V_{y}^{}\sG_{}^{R}V_{x}^{}\sG_{}^{A}
\big\rangle d\cE-\nonumber\\[4pt]
&\,-\frac{\hbar e_{}^{2}}{2\pi}\Re\!\int\! f(\cE)\Tr\Big\langle V_{y}^{}
\sG_{}^{A}V_{x}^{}\fpt{\sG_{}^{A}}{\cE}-(x\rightleftarrows y)\Big\rangle d\cE.
\end{align}
Here, $f(\cE)$ is the Fermi--Dirac function, $\sG_{}^{R(A)}(\cE)=1/(\cE-\cH-U
\pm i0)$ is the exact (i.e., non-averaged) retarded ($R$) or advanced ($A$) GF
operator of the Schr\"{o}dinger equation with total Hamiltonian $\cH+U$
(\ref{eq:chern_ham}), $V_{y(x)}^{}=v\sigma_{y(x)}^{}$ are Cartesian components
of the velocity operator, the symbol $\Tr=\tr\Sp$ denotes the trace over both
the spatial ($\Sp$) and (pseudo-) spin ($\tr$) degrees of freedom, the symbol
$(x\rightleftarrows y)$ denotes term that is obtained from the previous one due
to permutation $V_{x}^{}\rightleftarrows V_{y}^{}$, and, finally, angular
brackets $\langle\ldots\rangle$ denote averaging over the random field $U$
configurations. The normalization volume (area) is assumed to be unity.

The first term of conductivity (\ref{eq:kubo_total}), $\sigma_{H}^{\rm I}$,
results from the electrons at the Fermi surface, whereas $\sigma_{H}^{\rm II}$
is determined by the contribution of all occupied states of the Fermi sea.
St\v{r}eda et al.\cite{streda} were first to show that, for spinless electrons, 
this part of the conductivity is equal to
\begin{equation}
\label{eq:spinless_streda}
\sigma_{H}^{\rm II}=|e|c\left(\fpt{n}{B}\right)_{\zeta}^{},
\end{equation}
where $n$ is the electron concentration, $B$ is a magnetic field orthogonal to
the considered system ($||{\rm OZ}$), and $\zeta$ is the chemical potential of
the electron gas. It should be pointed out that Eq.~(\ref{eq:spinless_streda})
is exact, and with thermodynamic Maxwell relation $\sigma_{H}^{\rm II}$ can be
expressed through $\big(\partial M/\partial\zeta\big)_{B}^{}$, where $M$ is the
magnetization of the electron gas. Detailed discussion of $\sigma_{H}^{\rm II}$ 
and its physical interpretation can be found in Pruisken's 
survey\cite{pruisken_1990}. As shown in Ref.~\onlinecite{nov_grosh_2006}, the 
expression (\ref{eq:spinless_streda}) is converted to
\begin{equation}
\label{eq:true_streda}
\sigma_{H}^{\rm II}=|e|c\left[\left(\fpt{n}{B}\right)_{\zeta}^{}-\left(\fpt
{M_{p}^{}}{\zeta}\right)_{B}^{}\right]
\end{equation}
if the electron spin degree of freedom taken into account. Here, $M_{p}^{}$ is
the spin magnetization of the electron gas. It follows that in the general case,
$\sigma_{H}^{\rm II}$ is determined only by the orbital (diamagnetic) part of
the electron gas magnetization.

\section{\label{sec:green_func} One-electron Green's function and densities
of states}

The terms of the perturbation theory series for the electrical conductivity
(\ref{eq:kubo_total}) are expressed in terms of the averaged retarded
(advanced) GFs, which can be represented in the Dyson form
\begin{equation}
\label{eq:dyson_form}
G_{}^{R(A)}(\cE)=\big\langle\sG_{}^{R(A)}(\cE)\big\rangle=\frac{1}
{\cE-\cH-\Sigma_{}^{R(A)}(\cE)},
\end{equation}
where $\Sigma_{}^{R(A)}(\cE)$ is the electron self-energy operator. Below, in
specific calculations, we restrict ourselves to the SCBA in which
$\Sigma_{}^{R(A)}(\cE)=W\Sp G_{}^{R(A)}(\cE)$, where $W$ is the amplitude of
the pair correlator of the Gaussian random field $U$. It is easy to verify that, 
in this approximation, $\Sigma(\cE)$ is diagonal in spin space and has the
following matrix structure\cite{footnote_1}
\begin{equation}
\label{eq:self_en_matrix}
\Sigma=\Sigma_{e}^{}+\Sigma_{m}^{}\sigma_{z}^{},\quad~~\Sigma_{e}^{}=\frac{W}
{2}\Tr G,\quad~~\Sigma_{m}^{}=\frac{W}{2}\Tr\sigma_{z}^{}G.
\end{equation}
Hence, it follows that the averaged GF (\ref{eq:dyson_form}) can be obtained
from the GF of the clean model $G_{}^{0}=1/(\cE-\cH)$ using the
substitutions\cite{dug_etal_2005,nov_grosh_2006}
\begin{equation}
\label{eq:substitution}
\cE\mapsto\tilde\cE=\cE-\Sigma_{e}^{},\qquad M\mapsto\tM= M+\Sigma_{m}^{}.
\end{equation}
Here, $\Sigma_{e}^{}=\Delta_{e}^{}\mp i\hbar/2\tau_{e}^{}$ describes the
perturbation (shift $\Delta_{e}^{}$ and broadening $\hbar/\tau_{e}^{}$) of the
one-electron energy levels by a random field. The real part of the
$\Sigma_{m}^{}=\Delta_{m}^{}\mp i\hbar/2\tau_{m}^{}$ determines the
renormalization of the Dirac electron mass $M$, while its imaginary part
$\propto\hbar/\tau_{m}^{}$ makes a~contribution to the overall lifetime of the
one-electron states in the spin-split subbands. Thus, the one-particle GF 
operator averaged over random field configurations takes the following
form\cite{sin_etal_2006,sin_etal_2007,nov_2019}
\begin{equation}
\label{eq:gf_in_b}
G=\frac{\tilde\cE+\tM\sigma_{z}^{}+v(\vpi\cdot\vsigma)}
{\tilde{\cal E}_{}^{2}-\tilde{M}_{}^{2}-2mv_{}^{2}\cH_{0}^{}}=
\big[\tilde\cE+\tM\sigma_{z}^{}+v(\vpi\cdot\vsigma)\big]\cG,
\end{equation}
where
\begin{equation}
\label{eq:ideal_ham}
\cH_{0}^{}=\frac{1}{2m}(\vpi\cdot\vsigma)_{}^{2}=\frac{\vpi_{}^{2}}{2m}+\frac
{\hbar\omega_{c}^{}}{2}\sigma_{z}^{}
\end{equation}
is the Hamiltonian of a free electron with ideal value of Zeeman coupling
(g = 2) in an orthogonal magnetic field. In the limit $B\to 0$, $\vpi$ and
$\cH_{0}^{}$ in (\ref{eq:gf_in_b}), (\ref{eq:ideal_ham}) are replaced by
$\bm p$ and $\cE_{\bm p}^{}$, respectively. It should be emphasized that the 
second term in Hamiltonian $\cH_{0}^{}$ (\ref{eq:ideal_ham}) appears due to
commutation properties of the operators $\vpi$ and $\vsigma$ and has no
relation to the true Zeeman energy of an electron in a~magnetic field.

\begin{figure*}[t!]
\vspace*{-3.1cm}
\hspace*{-0.4cm}\includegraphics[scale=0.62,angle=-90]{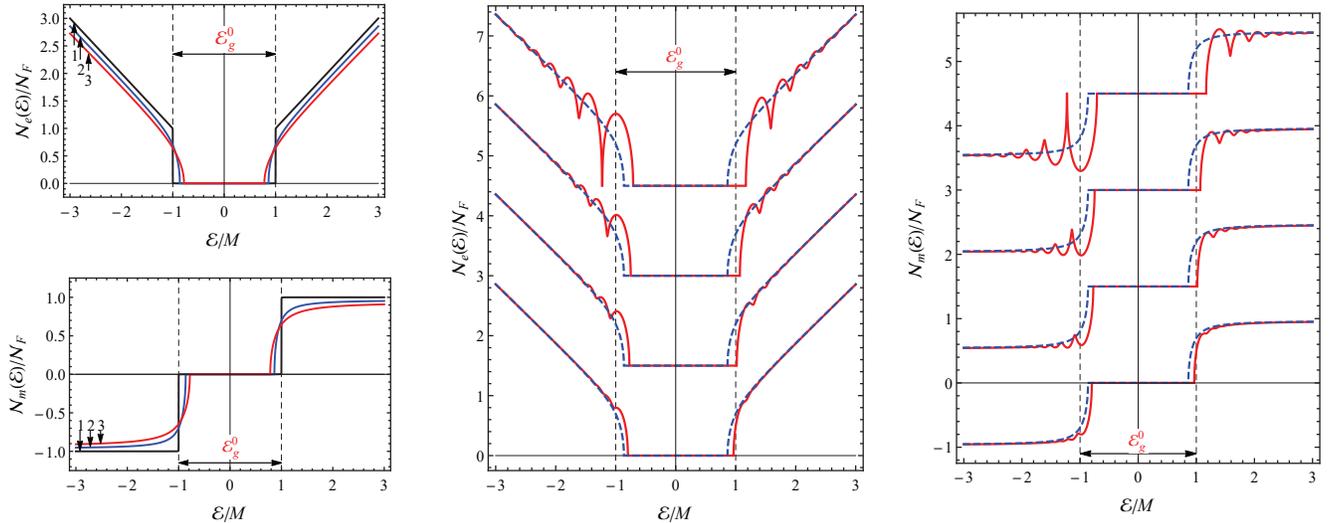}
\vspace*{-0.75cm}
\caption{\label{fig:dos_ddos} (color online) \underline{Left top panel}: The
energy dependence of the DOS $\cN_{e}^{}(\cE)=\cN_{\uparrow}^{}(\cE)+
\cN_{\downarrow}^{}(\cE)$ calculated in SCBA for various values of the disorder
parameter $\gamma_{0}^{}=W\cN_{F}^{}/2mv_{}^{2}$. Three curves are
calculated for $B=0$ and $\gamma_{0}^{}=0.0~~(1),~0.02~~(2), ~0.04~~(3)$.
Vertical dashed lines represent unperturbed edges of the electronic spectrum
$\cE=\pm|M|$. \\
\underline{Left bottom panel}: The same for the SDOS $\cN_{m}^{}(\cE)=
\cN_{\uparrow}^{}(\cE)-\cN_{\downarrow}^{}(\cE)$.\\
\underline{Central panel}: The energy dependence of the DOS calculated in SCBA
for $\gamma_{0}^{}=0.02$  and  $\hbar\omega_{c}^{}/M=0.50,\;0.33,\;0.25,\;0.17$
(from top to bottom). The dashed curves depict the energy dependence of DOS for
$\gamma_{0}^{}=0.02$ and $B=0$. For clarity, the curves are shifted relative to
each other along the abscissa axis.\\
\underline{Right panel}:  The same for the SDOS $\cN_{m}^{}(\cE)$.}
\vspace*{-0.3cm}
\end{figure*}

Eqs. (\ref{eq:self_en_matrix})  and (\ref{eq:gf_in_b}) form a system of the
self-con\-sis\-tent transcendental equations for the $\Sigma_{e}^{}$ and
$\Sigma_{m}^{}$. In the absence of an external magnetic field, the traces $\Tr
G$ and $\Tr\sigma_{z}^{}G$ are proportional to each other; therefore, both of
these self-energies in SCBA are expressed through the same
function\cite{nov_2019}
\begin{equation}
\label{eq:self_en_b0}
\Sigma_{e}^{}=\gamma_{0}^{}\cE\Phi,\quad\Sigma_{m}^{}=\gamma_{0}^{}M\Phi,
\quad|\gamma_{0}^{}\Phi|<<1
\end{equation}
that satisfies the equation
\begin{equation}
\label{eq:phi_func}
\Phi=\ln\left[(1+\gamma_{0}^{}\Phi)_{}^{2}-(1-\gamma_{0}^{}\Phi)_{}^{2}\frac
{{\cE}_{}^{2}}{M_{}^{2}}\right].
\end{equation}
Here $\gamma_{0}^{}=W\cN_{F}^{}/2mv_{}^{2}$ is the dimensionless parameter
of disorder, $\cN_{F}^{}=m/2\pi\hbar_{}^{2}$ is the density of states (DOS) of
two-dimensional free non-relativistic spinless electrons in the absence of
a~magnetic field. The numerical solution of equation (\ref{eq:phi_func}) allows 
us to calculate the self-energies $\Sigma_{e}^{}$ and $\Sigma_{m}^{}$, as well 
as the total DOS $\cN(\cE)$ and the difference of partial DOSs $\cN_{m}^{}
(\cE)$ with opposite values of the pseudospin projections onto the OZ-axis (SDOS).
\begin{equation}
\label{eq:dos_ddos}
\cN(\cE)=\frac{1}{\pi}\Im\Tr G_{}^{A}(\cE),\quad~~\cN_{m}^{}(\cE)=\frac{1}{\pi}
\Im\Tr\sigma_{z}^{}G_{}^{A}(\cE).
\end{equation}
Some results of this calculation at the different values of the parameter
$\gamma_{0}^{}$ are shown in Fig.~\ref{fig:dos_ddos}. As can be seen from
this figure, the energy gap between valence and conductivity bands narrows as
$\gamma_{0}^{}$ increases.

Now, we consider the calculation of the electron self-energies
(\ref{eq:self_en_matrix}) and DOSs (\ref{eq:dos_ddos}) in the case of $B\neq
0$. The spectrum of Hamiltonian $\cH_{0}^{}$ (\ref{eq:ideal_ham}) consists of 
two unbounded from above sets of the equidistant Landau levels $\cE_{n}^{}=
\hbar\omega_{c}^{}(n+1/2\pm 1/2)$, where $n = 0,1,2,\ldots$. As a~consequence,
the traces that determine the electronic self-energies (\ref{eq:self_en_matrix})
diverge logarithmically. These divergences are non-physical and can be
eliminated by the appropriate redefinition of the quantities $\cE$, $M$,
$\Sigma_{e}^{}$, and $\Sigma_{m}^{}$ (see, for example,
Ref.~\onlinecite{pruisken_1990}). As a~result, the traces $\Tr G$ and $\Tr
\sigma_{z}^{}G$ can be written as\cite{nov_2019}
\begin{equation}
\label{eq:tr_dtr_in_b}
\left[\begin{array}{c} \Tr G\\[2pt] \Tr\sigma_{z}^{}G \end{array}\right]=\frac
{\cN_{F}^{}}{mv_{}^{2}}\left\{\left[\begin{array}{c} \tilde{\cal E}\\ \tilde{M} 
\end{array}\right]\Phi(z)+\left[\begin{array}{c} \tilde{M}\\ \tilde{\cE} 
\end{array}\right]\frac{1}{2z}\right\},
\end{equation}
where
\begin{equation}
\label{eq:phi_in_b}
\Phi(z)=\psi(z)+\frac{1}{2z}+\ln\frac{2mv_{}^{2}\hbar\omega_{c}^{}}{M_{}^{2}},
\quad z=\frac{\tM_{}^{2}-\tilde{\cal E}_{}^{2}}{2mv_{}^{2}\hbar\omega_{c}^{}},
\end{equation}
and $\psi(z) = d/dz\ln\Gamma(z)$ is the digamma-function\cite{abram_steg_1964}.
The last term in the right-hand side of Eq.~(\ref{eq:phi_in_b}) eliminates the
divergence of the digamma-function as $B\to 0$ and provides the correct
passage of (\ref{eq:phi_in_b}) to this limit.

Some results of  the numerical solution of the SCBA-equations
(\ref{eq:self_en_matrix}), (\ref{eq:tr_dtr_in_b}), (\ref{eq:phi_in_b}) at the
different values of the magnetic induction are shown in Fig.~\ref{fig:dos_ddos}.
As can be seen from this figure, switching-on of an external magnetic field
leads to violation of the electron-hole symmetry due to the presence of the
anomalous Landau level $\cE_{-1,0}^{}=-M$ (\ref{eq:landau_levels}) at the top
of the valence band. This manifests itself in a~shift of the electronic
spectrum gap $\cE_{g}^{}$ towards higher energies compared to its location in
the absence of a~magnetic field. Finally, the de Haas--van Alphen type
oscillations in the energy dependence of the DOSs appear as the magnetic field 
$B$ increases.

\section{\label{sec:streda_term} Topological nature of St\v{r}eda-like term of 
conductance}

\begin{figure}[t!]
\vspace*{0.2cm}
\hspace*{-0.1cm}\includegraphics[scale=0.46]{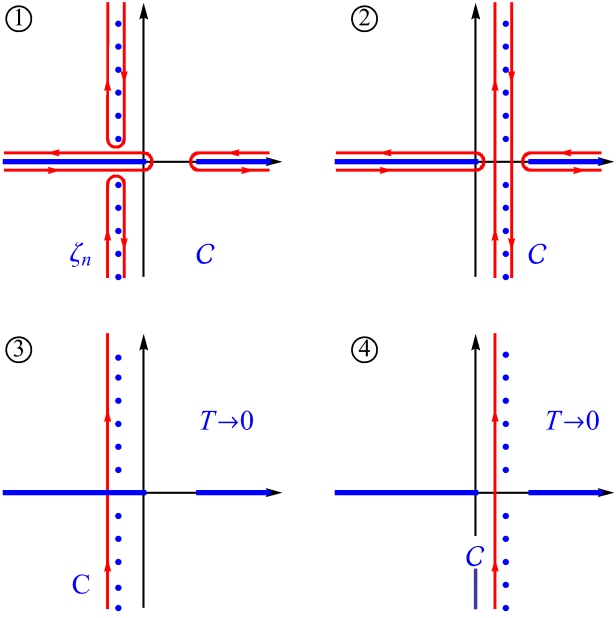}
\vspace*{0.3cm}
\caption{\label{fig:contours} (color online) Deformation of the integration
contour $C$. The bullets depict the poles of the Fermi--Dirac function
$\zeta_{n}^{}=\zeta+i\pi k_{B}^{}T(2n+1)$. Thick horizontal lines represent two
neighboring energy bands separated by a~gap $\cE_{g}^{}$.}
\end{figure}

Let us take St\v{r}eda-like formula (\ref{eq:true_streda}) as the initial
expression for calculating of the $\sigma_{H}^{\rm II}$ contribution to the
intrinsic anomalous Hall conductance. With this purpose, we have to find the
explicit expressions for the thermodynamic derivatives $\big(\partial n/
\partial B\big)_{\zeta}^{}$ and/or $\big(\partial M_{p}^{}/\partial\zeta
\big)_{B}^{}$ in the presence of an orthogonal magnetic field $B||{\rm OZ}$
[See Eq.~(\ref{eq:true_streda})]. For example, we consider the derivative
$\big(\partial n/\partial B\big)_{\zeta}^{}$. By definition, the electron
concentration is equal to
\begin{equation}
\label{eq:n_def}
n=\int\!f(\cE)\cN(\cE)\,d\cE=\frac{1}{2\pi i}\int_{C}f(z)\Tr G(z)\,dz.
\end{equation}
Here, $G(z)$ is the one-electron GF defined on the complex $z$--plane in such
a way that $\lim_{z\to{\cal E}\pm i0}G(z)=G_{}^{R(A)}(\cE)$; the integration
contour $C$ encircles counterclockwise the energy intervals belonging to the
spectrum of the eigenvalues of the Hamiltonian (\ref{eq:chern_ham}). Using the
analytical properties of the integrand in Eq.~(\ref{eq:n_def}), we can deform
$C$ as shown in Fig.~\ref{fig:contours}. This deformed integration contour
encircles clockwise the poles of the Fermi--Dirac function $\zeta_{n}^{}=\zeta+
i\pi k_{B}^{}T(2n+1)$ (See Figs.~\ref{fig:contours}.1 or~\ref{fig:contours}.2).
The integrals along the parts of $C$ located to the right of the poles
$\zeta_{n}^{}$ vanish as $T\to 0$. In this case, the integration contour in
Eq.~(\ref{eq:n_def}) looks as shown in Figs.~\ref{fig:contours}.3
or~\ref{fig:contours}.4.

Thus, the problem is reduced to finding of an explicit expression for
derivative $\partial\Tr G/\partial B$. This can be done in the momentum
representation in which Dyson's equation for the averaged one-particle GF
has the form\cite{onoda_etal_2008}
\begin{equation}
\label{eq:dyson_p_repres}
Q_{\bm p}^{}\star G_{\bm p}^{}=\left(z-\cH_{\bm p}^{}-\frac{\hbar\omega_{c}^{}}
{4}s_{z}^{}\otimes g-\Sigma_{\bm p}^{}\right)\star G_{\bm p}^{}=1.
\end{equation}
Here, $G_{\bm p}^{}$ is the Fourier transform of a gauge-independent and
translation invariant multiplier of the averaged one-particle GF in coordinate
representation $G(\bm r,\bm r')=\aver{\bm r|G|\bm r'}$ (See 
Appendix~\ref{app:moyal}), $\cH_{\bm p}^{}$ is the Hamiltonian of a~some clean
non-interacting system in which the momentum operator $\vpi$ is replaced by
$c$-number momentum $\bm p$ (further, we will omit subscript $\bm p$). For
clarity, the Zeeman energy $\hbar\omega_{c}^{}s_{z}^{}\otimes g/4$ is written
explicitly in (\ref{eq:dyson_p_repres}), where $s_{z}^{}$ is the Pauli spin
$z$-matrix, $g$ is the matrix of the effective $g$-factors. For example, $g=
{\rm diag}\{g_{e}^{},g_{h}^{}\}$ in the Bernevig--Hughes--Zhang model for the
single valley Dirac fermions in the HgTe quantum well
\cite{bern_etal_2006,buet_etal_2011}.

Symbol $F_{\bm p}^{}\star G_{\bm p}^{}$ denotes the Moyal $\star$-product
\cite{onoda_etal_2008,moyal_1949,zach_etal_2005} of the momentum-dependent
functions
\begin{equation}
\label{eq:star_product}
F_{\bm p}^{}\star G_{\bm p}^{}\stackrel{\rm def}{=}F_{\bm p}^{}\exp\left[-i
\frac{\hbar_{}^{2}}{2l_{B}^{2}}\epsilon_{\alpha\beta}^{}
\stackrel{\leftarrow}{\partial}_{\alpha}^{}
\stackrel{\rightarrow}{\partial}_{\beta}^{}\right]G_{\bm p}^{},
\end{equation}
where $l_{B}^{}=\sqrt{c\hbar/|e|B}$ is the magnetic length,
$\epsilon_{\alpha\beta}^{}$ is the antisymmetric unit tensor of the second rank
[$\alpha\,(\beta)=x,\,y$], and $\partial_{\alpha}=\partial/\partial
p_{\alpha}^{}$. The arrows above the differentiation operators indicate the
directions of their actions. The formulation of the momentum representation for
the description of electron states in a~magnetic field and an overview of the
various definitions and basic properties of the Moyal $\star$-product can be
found in Appendix~\ref{app:moyal}.

In the momentum representation, the operations $\Tr$ and $\partial/\partial B$
commute with each other. So, we just need to find an explicit expression for
$\partial G/\partial B$ to calculate $\big(\partial n/\partial B)_{\zeta}^{}$.
Direct differentiation of Dyson equation (\ref{eq:dyson_p_repres}) with respect
to the magnetic field induction gives the following result
\begin{align}
\label{eq:dg_db_1}
\fpt{G}{B}=&-G\star\fpt{Q}{B}\star G+i\frac{\hbar|e|}{2c}
\epsilon_{\alpha\beta}^{}G\star\fpt{Q}{p_{\alpha}^{}}\star\fpt{G}{p_{\beta}^{}}
\nonumber\\
=&-G\star\fpt{Q}{B}\star G-i\frac{\hbar|e|}{2c}\epsilon_{\alpha\beta}^{}G\star
\fpt{Q}{p_{\alpha}^{}}\star G\star\fpt{Q}{p_{\beta}^{}}\star G.
\end{align}
Here, the last equality is obtained using the first of the identities
\begin{equation}
\label{eq:other_deriv}
\fpt{G}{\bm p}=-G\star\fpt{Q}{\bm p}\star G,\qquad\fpt{G}{z}=-G\star\fpt{Q}
{z}\star G,
\end{equation}
which are also derived from Dyson equation (\ref{eq:dyson_p_repres}).

Let us take a trace of Eq.~(\ref{eq:dg_db_1}). Then, using the second identity
from (\ref{eq:other_deriv}) and the trace invariance under cyclic permutations,
we obtain
\begin{align}
\label{eq:tr_dg_db_1}
\Tr\fpt{G}{B}=&\Tr\fpt{Q}{B}\star\fpt{G}{z}-\Tr G\star\fpt{Q}{B}\star G\star
\fpt{\Sigma}{z}-\nonumber\\
&\hspace*{0.5cm}-i\frac{\hbar|e|}{2c}\epsilon_{\alpha\beta}^{}\Tr G\star\fpt{Q}
{p_{\alpha}^{}}\star G\star\fpt{Q}{p_{\beta}^{}}\star G.
\end{align}
Now we substitute the explicit expression
\begin{equation}
\label{eq:dq_db}
\fpt{Q}{B}=-\frac{\hbar|e|}{4mc}s_{z}^{}\otimes g-\fpt{\Sigma}{B}
\end{equation}
in the first term of the right-hand side of Eq.~(\ref{eq:tr_dg_db_1}) and 
transfer the Zeeman term to the left-hand side of this equality. The electron 
self energy satisfies relations like Ward identity
\begin{equation}
\label{eq:ward_s}
\fpt{\Sigma_{n}^{}}{z}=\sum_{n_{}'}U_{n,n_{}'}^{}\fpt{G_{n_{}'}^{}}{z},
\quad~~\fpt{\Sigma_{n}^{}}{B}=\sum_{n_{}'}U_{n,n_{}'}^{}\fpt{G_{n_{}'}^{}}{B},
\end{equation}
where $U_{n,n_{}'}^{}$ is the two-particle irreducible interaction vertex.
Hence, it follows that
\begin{equation}
\label{eq:ward_1}
\Tr\fpt{\Sigma}{B}\star\fpt{G}{z}=\Tr\fpt{G}{B}\star\fpt{\Sigma}{z}.
\end{equation}
This is more strong version of the relation previously proposed by
Pruisken\cite{pruisken_1990}. With this in mind Eq.~(\ref{eq:tr_dg_db_1}) can
be rewritten as
\begin{align}
\label{eq:trace_dg_db}
\Tr\fpt{G}{B}+&\frac{\hbar|e|}{4mc}\Tr s_{z}^{}\otimes g\fpt{G}{z}=-i\frac
{\hbar|e|}{2c}\epsilon_{\alpha\beta}^{}\times\nonumber\\
&\hspace*{1.0cm}\times\Tr\fpt{Q}{p_{\alpha}^{}}\star G\star\fpt{Q}{p_{\beta}^{}}
\star G\star\fpt{Q}{z}\star G.
\end{align}

After integration with respect to  $z$ of this equation with Fermi--Dirac
weight function $f(z)$, the terms of its left-hand side determine derivatives
of the electron concentration and spin magnetization included in
Eq.~(\ref{eq:true_streda}). Therefore, after some cosmetic transformations, the
expression for the St\v{r}eda-like term (\ref{eq:true_streda}) of the Hall
conductance can be written in the following form
\begin{align}
\label{eq:aver_streda}
\sigma_{H}^{\rm II}=&-\frac{\hbar e_{}^{2}}{12\pi}\!\int_{C}\!f(z)\times
\nonumber\\[4pt]
&\times\epsilon_{\alpha\beta\gamma}^{}\Tr\fpt{Q}{p_{\alpha}^{}}\star G\star
\fpt{Q}{p_{\beta}^{}}\star G\star\fpt{Q}{p_{\gamma}^{}}\star G\,dz.
\end{align}
Here, integration is carried out along the contour $C$ shown in
Fig.~\ref{fig:contours}.1 or Fig.~\ref{fig:contours}.2; 
$\epsilon_{\alpha\beta\gamma}^{}$ is the unit antisymmetric pseudotensor
of the third rank, indices $\alpha,\,\beta,\,\gamma$ run values 0,\,1,\,2 with
$p_{0}^{}=z$, $p_{1}^{}=p_{x}^{}$, and~$p_{2}^{}=p_{y}^{}$.

Thus, we obtain the expression for the St\v{r}eda-like part of the Hall
conductance in which the averaging over random field configurations is
performed exactly. Eq.~(\ref{eq:aver_streda}) is true regardless of the
disorder type and the approximation for the averaged one-particle GF. This is
possible owing to special analytical structure of starting expression for
$\sigma_{H}^{\rm II}$ (\ref{eq:kubo_total}) that does not contain the products
$\sG_{}^{R}\sG_{}^{A}$. It should be emphasized that both expressions for
$\sigma_{H}^{\rm II}$ (\ref{eq:kubo_total}) and (\ref{eq:aver_streda}) have the
identical structure. The only difference is that Eqs. (\ref{eq:kubo_total}) and
(\ref{eq:aver_streda}) include exact and averaged GFs, respectively. In other
words, the averaging procedure of $\sigma_{H}^{\rm II}$ is reduced to the
simple substitute the exact GFs $\sG$ for the averaged ones $G=\aver\sG$.
If desired, Eq.~(\ref{eq:aver_streda}) can be rewritten in the usual operator
representation. To do this, it is enough to replace the $\star$-products in
this equation by the usual multiplication and treat $G$ and $Q$ as functions
depending on the operators of the corresponding dynamic variables.

Eq.~(\ref{eq:aver_streda}) indicates a close connection between the intrinsic
anomalous Hall conductance and topological properties of the one-electron
states. Indeed, in the case $T = 0$, the Fermi--Dirac function $f(z)$ coincides
with the Heaviside unit step $\Theta(\cE_{F}^{}-\Re z)$, and  the expression
for $\sigma_{H}^{\rm II}$ (\ref{eq:aver_streda}) takes the form
\begin{align}
\label{eq:streda_chern_1}
\sigma_{H}^{\rm II}&\,=\sigma_{0}^{}\!\int_{C}\!\left\{-
\epsilon_{\alpha\beta\gamma}^{}\frac{1}{24\pi_{}^{2}}\!\iint\times\right.
\nonumber\\[6pt]
&\,\left.\times\tr\left[G\star\fpt{Q}{p_{\alpha}^{}}\star G\star\fpt{Q}
{p_{\beta}^{}}\star G\star\fpt{Q}{p_{\gamma}^{}}\right]dp_{1}^{}\,dp_{2}^{}
\right\}dp_{0}^{},
\end{align}
where the integration contour $C$ is shown in Figs.~\ref{fig:contours}.3
or~\ref{fig:contours}.4 and trace over the spatial degrees of freedom is 
presented as the integral with respect to momenta  $p_{1,2}^{}$ 
(\ref{eq:trace_formula_s}).

Now imagine that the Fermi level lies inside the energy gap of the system under
consideration. In this case, the integration contour $C$ in
Eq.~(\ref{eq:streda_chern_1}) is the straight line parallel to the imaginary
axis $\Re z=\cE_{F}^{}-0$, $-\infty< \Im z<+\infty$ (See
Fig.~\ref{fig:contours}.4) and the St\v{r}eda term of the Hall conductance 
$\sigma_{H}^{\rm II}$ takes the disorder-independent quantized value 
proportional to the Chern number, i.e., $\sigma_{H}^{\rm II}=\sigma_{0}^{}{\rm 
Ch}$, where
\begin{equation}
\label{eq:chern_num_gen}
{\rm Ch}=-\frac{\epsilon_{\alpha\beta\gamma}^{}}{24\pi_{}^{2}}\!\iiint\!\tr\!
\left[G\star\fpt{Q}{p_{\alpha}^{}}\star G\star\fpt{Q}{p_{\beta}^{}}\star G
\star\fpt{Q}{p_{\gamma}^{}}\right]\!d_{}^{3}p.
\end{equation}
Integration is performed here over the entire $(2+1)$-dimensional space of 
points $(\Im z,p_{x}^{},p_{y}^{})$. Eq. (\ref{eq:chern_num_gen}) is an analogue
of the expression for the Chern number obtained by Zubkov et
al.\cite{zubkov_2016,zub_wu_2020,zhang_zub_2019} in the Wigner representation,
which is necessary for describing spatially nonuniform systems. Writing the 
Chern number in the momentum representation (\ref{eq:chern_num_gen}) is more 
appropriate in the practically important spatially uniform case.

We represent the result of integration along the straight line $\Re z=
\cE_{F}^{}-0$ in Eq.~(\ref{eq:chern_num_gen}) in the form of increment of the   
corresponding antiderivative, i.e.   
\begin{equation}
\label{eq:new_chern}
{\rm Ch}=\Delta F(\cE_{F}^{}\pm i\infty)=F(\cE_{F}^{}+i\infty)-F(\cE_{F}^{}-
i\infty),
\end{equation}
where [See Eq.~(\ref{eq:streda_chern_1})]
\begin{equation}
\label{eq:antideriv}
\fpt{F}{z}=-\epsilon_{\alpha\beta\gamma}^{}\!\iint\!\tr\!\left[G\star\fpt{Q}
{p_{\alpha}^{}}\star G\star\fpt{Q}{p_{\beta}^{}}\star G\star\fpt{Q}
{p_{\gamma}^{}}\right]\!\frac{d_{}^{2}\bm p}{24\pi_{}^{2}}.
\end{equation}
Obviously, the asymptotic values $F(\cE_{F}^{}\pm i\infty)$ do not depend on
the disorder and coincide with their values in the clean limit.

If the Fermi level lies outside the gap, then the contour of integration in
Eq.~(\ref{eq:streda_chern_1}) consists of two semi-infinite straight lines
parallel to the imaginary axis $\Re z=\cE_{F}^{}-0$, $-\infty< \Im z<-0$ and
$+0< \Im z<+\infty$ (See Fig.~\ref{fig:contours}.3). In this case, a~term
proportional to the discontinuity of $F(z)$ across the cut line corresponding
to the continuous spectrum of the electron is added to the expression for the
St\v{r}eda part of the Hall conductance. So, we have
\begin{equation}
\label{eq:streda_fin_top}
\sigma_{H}^{\rm II}=\sigma_{0}^{}\left\{\begin{array}{ll}
{\rm Ch}, & \cE_{F}^{}\in{\rm Gap},\\[4pt]
{\rm Ch}+\Delta F(\cE_{F}^{}), & \cE_{F}^{}\notin{\rm Gap},\end{array}\right.
\end{equation}
where $\Delta F(\cE_{F}^{})=F(\cE_{F}^{}-i0)-F(\cE_{F}^{}+i0)$. The first term
in the bottom line of (\ref{eq:streda_fin_top}) remains unchanged since $\Delta
F(\cE_{F}^{}\pm i\infty)$ does not depend on placing the Fermi level. In the
next section, we will illustrate the validity of (\ref{eq:streda_fin_top}) 
using the massive Dirac electrons model as an example.

\section{\label{sec:intr_hall} Intrinsic anomalous Hall conductance in SCBA}

The intrinsic Hall conductance consists of two terms $\sigma_{H}^{\rm int}=
\sigma_{H}^{\rm Ib}+\sigma_{H}^{\rm II}$, where $\sigma_{H}^{\rm Ib}$ is the
bare bubble part of $\sigma_{H}^{\rm I}$ and $\sigma_{H}^{\rm II}$ is the
St\v{r}eda term [see. Eq.~(\ref{eq:kubo_total})].

In the presence of an external magnetic field, the Hall conductance consists of
the normal $\sigma_{H}^{\rm n}$ and anomalous $\sigma_{H}^{\rm a}$ terms, at
that $\sigma_{H}^{\rm n}(-B)=-\sigma_{H}^{\rm n}(B)$ and $\sigma_{H}^{\rm a}
(-B)=\sigma_{H}^{\rm a}(B)$. Below, we restrict ourselves to calculating the
even in $B$ parts of $\sigma_{H}^{\rm Ib}$ and $\sigma_{H}^{\rm II}$. Of course,
the parity of the anomalous Hall conductance in the magnetic field does not
mean violation of the Onsager relation. The point is that this relation is
valid for the both normal and anomalous Hall conductances with simultaneous
reversal of the signs of both the magnetic field $B$ and the Dirac mass $M$, 
since the introduction of each of them breaks the time reversal symmetry. Thus, 
the Onsager relation for both parts of $\sigma_{H}^{}$ must have the form
\begin{equation}
\label{eq:onsager}
\sigma_{H}^{}(-B,-M)=-\sigma_{H}^{}(B,M).
\end{equation}
The intrinsic anomalous Hall conductance calculated below satisfies the
relation (\ref{eq:onsager}) not only in the QAHE
regime\cite{boet_etal_2019,boet_etal_2020}, but also outside it.

\subsection{\label{sec:intr_hall_cond} St\v{r}eda like part of the anomalous
Hall conductance}

It is usually assumed that $\sigma_{H}^{\rm II}$ is weakly dependent on
disorder, and it is calculated without taking into account the scattering of
charge carriers\cite{dug_etal_2005,sin_etal_2006,sin_etal_2007,nun_etal_2007,
ado_etal_2015,ado_etal_2016}. It will be shown below that this approximation is 
violated in a fairly wide vicinity of the Hall plateau. Let us rewrite the 
expression (\ref{eq:streda_chern_1}) for the St\v{r}eda term in a~more 
convenient for calculating operator notation
\begin{equation}
\label{eq:scba_streda_term}
\sigma_{H}^{\rm II}=\sigma_{0}^{}\frac{\hbar_{}^{2}v_{}^{2}}{2}\!\int_{C}^{}
\!\!\Tr\!\left[\sigma_{y}^{}G\sigma_{x}^{}\fpt{G}{z}-(x\leftrightarrows y)
\right]\!dz.
\end{equation}
Here, it is taken into account that $\partial Q/\partial p_{x(y)}^{}=-v
\sigma_{x(y)}^{}$ in SCBA considered below. For simplicity, we assume the
temperature to be zero, the generalization to the case $T>0$ is obvious.

In the article of Dugaev et al.\cite{dug_etal_2005}, the St\v{r}eda term of
$\sigma_{H}^{}$ was calculated for the case of a~clean two-dimensional Rashba
model in absence of a magnetic field. We will apply their approach to
calculate the St\v{r}eda part of the anomalous Hall conductance of the
disordered Chern insulator (\ref{eq:chern_ham}) in SCBA in the presence of
an~orthogonal magnetic field. Direct differentiation of the averaged GF
(\ref{eq:gf_in_b}) with respect to $z$ gives the following result
\begin{equation}
\label{eq:gf_derivat}
\fpt{G}{z}=\left(1-\fpt{\Sigma_{e}^{}}{z}+\fpt{\Sigma_{m}^{}}{z}\sigma_{z}^{}
\right)\cG-G\cG\fp{z}(\tilde{z}_{}^{2}-\tilde{M}_{}^{2}),
\end{equation}
where $\tilde{z}=z-\Sigma_{e}^{}(z)$. In absence of a~magnetic field, the
second term in this expression does not contribution to the Hall conductance
\cite{dug_etal_2005}. In the presence of a~magnetic field, this is no longer
the case, but only the first term from Eq.~(\ref{eq:gf_derivat}) makes an even
in $B$ contribution to the St\v{r}eda part of the Hall conductance. The
substitution of this term into the integrand of (\ref{eq:scba_streda_term})
gives, after some simple algebra, the following result
\begin{align}
\label{eq:first_contrib}
\Tr[\ldots]=&\,\frac{i}{mv_{}^{2}\hbar\omega_{c}^{}}\times\nonumber\\
&\,\times\Tr(\bar\cG-\cG)(\tilde{z}-\tM\sigma_{z}^{})\left(1-\fpt{\Sigma_{e}^{}}
{z}+\fpt{\Sigma_{m}^{}}{z}\sigma_{z}^{}\right),\nonumber
\intertext{where $\bar\cG=\sigma_{x(y)}^{}\cG\sigma_{x(y)}^{}$, i.e., this is
the result of transformation $\sigma_{z}^{}\mapsto-\sigma_{z}^{}$}
=&\,2i\frac{\cN_{F}^{}}{mv_{}^{2}}\frac{\tM\left(1-\fpt{\Sigma_{e}^{}}{z}\right)
-\tilde{z}\fpt{\Sigma_{m}^{}}{z}}{\tilde{z}_{}^{2}-\tilde{M}_{}^{2}}
\nonumber\\
=&\,i\frac{\cN_{F}^{}}{mv_{}^{2}}\fp{z}\ln\frac{\tilde{M}-\tilde{z}}
{\tilde{M}+\tilde{z}}.
\end{align}
Thus, we obtain the  explicit expression for the antiderivative [see
Eq.~(\ref{eq:antideriv})]
\begin{equation}
\label{eq:antideriv_form}
F(z)=\frac{1}{4\pi i}\ln\frac{\tilde{M}+\tilde{z}}{\tilde{M}-\tilde{z}}
\end{equation}
in the case of the model of disordered massive Dirac electrons. Obviously, its 
increment along the straight line $\Re z=\cE_{F}^{}$ does not depend on the 
placing of the Fermi level $\cE_{F}^{}$, nor on the external magnetic field $B$, 
nor on the electronic self-energies $\Sigma_{e(m)}^{}$, and is equal to the 
Chern number $\Delta F(\cE_{F}^{}\pm i\infty)=1/2={\rm Ch}$ 
(\ref{eq:new_chern}).

If $\cE_{F}^{}\in{\rm Gap}$, i.e., when $\Im\Sigma_{e(m)}^{}=0$, then $F(z)$  
is continuous across the real axis. As a~result, the anomalous part of the
St\v{r}eda term takes on quantized value $\sigma_{H}^{\rm II\, a}=\sigma_{0}^{}
{\rm Ch}$ in this energy range [see Eq.~(\ref{eq:streda_fin_top})]. Otherwise
($\cE_{F}^{}\notin{\rm Gap}$), $\sigma_{H}^{\rm II\,a}$ acquires additional
term $\propto\,\Delta F(\cE_{F}^{})=F(\cE_{F}^{}-i0)-F(\cE_{F}^{}+i0)$ [see 
Eq.~(\ref{eq:streda_fin_top})]. Given the explicit form of the function $F(z)$ 
(\ref{eq:antideriv_form}), we obtain the following expression for the
St\v{r}eda term of the anomalous Hall conductance
\begin{equation}
\label{eq:streda_term_b_0}
\sigma_{H}^{\rm II\,a}=\frac{\sigma_{0}^{}}{2}\big[1+2\Delta F(\cE_{F}^{})\big]
=\frac{\sigma_{0}^{}}{2\pi}\Im\ln\frac{\tilde{\cal E}_{}^{A}+\tilde{M}_{}^{A}}
{\tilde{\cal E}_{}^{A}-\tilde{M}_{}^{A}},
\end{equation}
\begin{figure*}[t!]
\vspace*{-2.5cm}
\hspace*{-0.4cm}\includegraphics[scale=0.62,angle=-90]{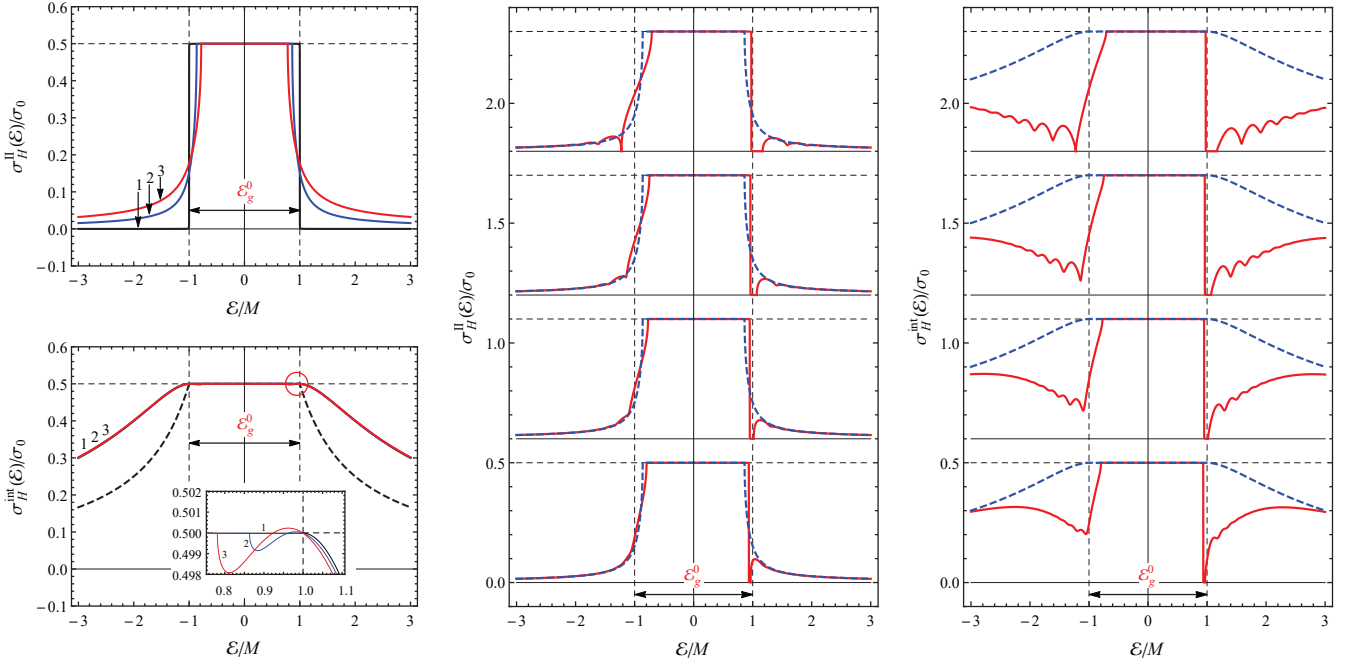}
\vspace*{-0.75cm}
\caption{\label{fig:intr_ahe_chern} (color online) \underline{Left top panel}:
The energy dependence of the St\v{r}eda term of the anomalous Hall conductance
$\sigma_{H}^{\rm II\,a}$ (\ref{eq:streda_term_b_0}) calculated in SCBA for
various values of the disorder parameter $\gamma_{0}^{}=W\cN_{F}^{}/2mv_{}^{2}$.
Three curves are calculated for $B=0$ and $\gamma_{0}^{}=0.0~~(1),~0.02~~(2),
~0.04~~(3)$. Vertical dashed lines represent unperturbed edges of the
electronic spectrum $\cE=\pm|M|$. \\
\underline{Left bottom panel}: The same for the total intrinsic part of the
anomalous Hall conductance $\sigma_{H}^{\rm int}=\sigma_{H}^{\rm Ib\,a}+
\sigma_{H}^{\rm II\,a}$. The dashed line shows the energy dependence of
$\sigma_{H}^{\rm int}=\sigma_{0}^{}|M|/2|\cE|$ of the clean ($\gamma_{0}^{}=0$) 
model of massive Dirac fermions for $B=0$. The solid curves calculated for
$B=0$ and $\gamma_{0}^{}\to 0~~(1),~=0.02~~(2),~=0.04~~(3)$ are practically 
indistinguishable at this scale. The inset shows, on an enlarged scale, the 
behavior of curves (1-3) in vicinity of the point $\cE=|M|$ (circled in the 
main graph).\\
\underline{Central panel}: The energy dependence of the St\v{r}eda term of the
anomalous Hall conductance $\sigma_{H}^{\rm II\,a}$ (\ref{eq:streda_term_b_0})
calculated in SCBA for $\gamma_{0}^{}=0.02$  and  $\hbar\omega_{c}^{}/M=0.50,\;
0.33,\;0.25,\;0.17$ (from top to bottom). The dashed curves depict the energy
dependence of $\sigma_{H}^{\rm II}$ for $\gamma_{0}^{}=0.02$ and $B=0$.
For clarity, the curves are shifted relative to each other along the abscissa
axis.\\
\underline{Right panel}: The same for the total intrinsic part of the anomalous
Hall conductance $\sigma_{H}^{\rm int}=\sigma_{H}^{\rm Ib\,a}+
\sigma_{H}^{\rm II\,a}$.}
\end{figure*}

\noindent which turns out to be valid both in the quantization regime and
beyond it. This expression for $\sigma_{H}^{\rm II\,a}$ has exactly the same
form as in absence of an external magnetic field\cite{nov_2019}. The entire
dependence on the magnetic field of (\ref{eq:streda_term_b_0}) is contained 
in the electronic self-energies $\Sigma_{e}^{}$ and $\Sigma_{m}^{}$
(\ref{eq:self_en_matrix}). The results of numerical analysis of
Eq.~(\ref{eq:streda_term_b_0}) are represent in  Fig.~\ref{fig:intr_ahe_chern}
(upper left and central panels).

\subsection{\label{sec:bare_sigma} Bare bubble part of the anomalous Hall
conductance}

Now we turn to the calculation of the bare bubble contribution
$\sigma_{H}^{\rm Ib\,a}$ to the intrinsic anomalous Hall conductance. 
In the case of the clean [$U(\bm r)=0$] massive Dirac electrons, this term 
was calculated by Sinitsyn and co-workers\cite{sin_etal_2007} in the
basis of the Hamiltonian $\cH$ (\ref{eq:chern_ham}) in absence of a~magnetic 
field. Here, we apply the simple algebraic approach to calculating
$\sigma_{H}^{\rm Ib\,a}$ of the considered model in the same approximation as
the St\v{r}eda term, that is, in SCBA.

Thus, we start with the expression
\begin{equation}
\label{eq:bare_bubble}
\sigma_{H}^{\rm Ib}=\frac{\hbar e_{}^{2}v_{}^{2}}{2\pi}\Tr\sigma_{y}^{}G_{}^{R}
\sigma_{x}^{}G_{}^{A}.
\end{equation}
Given the properties of the Pauli matrices, it can be shown that the terms
$\propto(\vpi\cdot\vsigma)$ in the numenator of GF (\ref{eq:gf_in_b}) do not 
contribute to $\Tr\sigma_{y}^{}G_{}^{R}\sigma_{x}G_{}^{A}$ due to isotropy of 
the considered system. So, substitution of the one-particle GFs 
(\ref{eq:gf_in_b}), (\ref{eq:ideal_ham}) in Eq.~(\ref{eq:bare_bubble}) gives
the result
\begin{align*}
\sigma_{H}^{\rm Ib}=&\,\frac{\hbar e_{}^{2}v_{}^{2}}{2\pi}\Tr\sigma_{y}^{}
\cG_{}^{R}(\tilde{\cal E}_{}^{R}+\tilde{M}_{}^{R}\sigma_{z}^{})\sigma_{x}^{}
(\tilde{\cal E}_{}^{A}+\tilde{M}_{}^{A}\sigma_{z}^{})\cG_{}^{A}\\
=&\,\frac{\hbar e_{}^{2}v_{}^{2}}{2\pi i}\Tr\sigma_{z}^{}(\tilde{\cal E}_{}^{R}
-\tilde{M}_{}^{R}\sigma_{z}^{})(\tilde{\cal E}_{}^{A}+\tilde{M}_{}^{A}
\sigma_{z}^{})\bar{\cal G}_{}^{R}\cG_{}^{A}.
\end{align*}
We are interested in the even in $B$ part of this expression
\begin{equation}
\label{eq:bare_calc}
\sigma_{H}^{\rm Ib\,a}=\frac{\hbar e_{}^{2}v_{}^{2}}{\pi}\Im
(\tilde{\cal E}_{}^{R}\tilde{M}_{}^{A})\Tr\bar{\cal G}_{}^{R}\cG_{}^{A}.
\end{equation}
Using the resolvent identity we write the product $\bar{\cal G}_{}^{R}
\cG_{}^{A}$ in the following form
\begin{align}
\label{eq:resolv_ident}
\bar{\cal G}_{}^{R}\cG_{}^{A}=\frac{1}{i\hbar}\frac{\tau}{2mv_{}^{2}}\frac
{1+i\omega_{c}^{}\tau\sigma_{z}^{}}{1+\omega_{c}^{2}\tau_{}^{2}}(\cG_{}^{A}-
\bar{\cal G}_{}^{R}).
\end{align}
The quantity $\tau$ that is defined by relation
\begin{equation}
\label{eq:tau_tr_def}
\frac{1}{\tau}=\frac{1}{mv_{}^{2}}\left(\frac{\cE}{\tau_{e}^{}}+\frac{M}
{\tau_{m}^{}}\right)
\end{equation}
plays the role of the transport time in the Drude-like denominator $1+
\omega_{c}^{2}\tau_{}^{2}$. Let us introduce one more characteristic time
\begin{equation}
\label{eq:tau_s_def}
\frac{1}{\tau'}=\frac{2}{\hbar}\frac{\Im({\tilde{\cal E}}_{}^{R}
{\tilde M}_{}^{A})}{mv_{}^{2}}=\frac{1}{mv_{}^{2}}\left(\frac{\cE}{\tau_{m}^{}}
+\frac{M}{\tau_{e}^{}}\right). 
\end{equation}
Substitution of expressions (\ref{eq:resolv_ident})--(\ref{eq:tau_s_def}) into 
Eq.~(\ref{eq:bare_calc}) gives the following result
\begin{equation}
\label{eq:bare_calc_fin}
\sigma_{H}^{\rm Ib\,a}=\frac{\sigma_{0}^{}}{2\pi}\frac{\tau}{\tau'}\frac
{\Im\Phi_{}^{A}+\ds\omega_{c}^{}\tau\Re\frac{mv_{}^{2}\hbar\omega_{c}^{}}
{\tilde{M}_{}^{2}-\tilde{\cal E}_{}^{2}}}{1+\omega_{c}^{2}\tau_{}^{2}},
\end{equation}
where the function $\Phi_{}^{A}$ is defined in (\ref{eq:phi_in_b}).

Obviously, expression (\ref{eq:bare_calc_fin}) as a function of $\cE_{F}^{}$
vanishes inside the gap of the one-electron energy spectrum. Thus, the 
quantization of the $\sigma_{H}^{\rm int}=\sigma_{H}^{\rm Ib\,a}+
\sigma_{H}^{\rm II\,a}$ (\ref{eq:streda_term_b_0}), (\ref{eq:bare_calc_fin})
survives when an external magnetic field is turned on that is consistent with 
recent results by B\"{o}ttcher et al.\cite{boet_etal_2019,boet_etal_2020}. The
results of numerical analysis of $\sigma_{H}^{\rm int}$ are represented in
Fig.~\ref{fig:intr_ahe_chern} (bottom left and right panels).

\section{\label{sec:discus} Results and discussion}

Let us summarize briefly the main results obtained in this work. The explicit 
expressions in SCBA (\ref{eq:tr_dtr_in_b}) are obtained for the total DOS and 
SDOS of the minimal model of a~two-dimensional disordered Chern insulator. The
results of a~numerical analysis of these expressions are shown in
Fig.~\ref{fig:dos_ddos}. In the case $B=0$, the curves in the left panels 
demonstrate a~narrowing of the gap in the spectrum of one-electron states as 
the disorder parameter increases. The graphs presented in central and right 
panels in Fig.~\ref{fig:dos_ddos} show two effects caused by an external 
magnetic field. First, the gap in the electronic spectrum is shifted towards 
higher energies as a~magnetic field increases. This is due to the presence of
the anomalous Landau level $\cE_{-1,0}^{}=-M$ at the top of the valence band.
Second, in the region of sufficiently strong magnetic fields, the de Haas--van  
Alphen oscillations manifest themselves in the energy dependencies of the DOSs.

Section \ref{sec:streda_term} provides the simple derivation of expressions
[see Eq.~(\ref{eq:aver_streda}) for the case $T>0$ and 
Eq.~(\ref{eq:streda_chern_1}) for the case $T=0$] for the St\v{r}eda-like term
$\sigma_{H}^{\rm II}$ (\ref{eq:true_streda}) of the anomalous Hall conductance 
in terms of the averaged one-particle GFs in the momentum representation. These
expressions are valid both in absence of an external magnetic field and in
its presence and describe the dependence of $\sigma_{H}^{\rm II}$ on the 
location of the Fermi level $\cE_{F}^{}$ both inside the gap of the
one-electron spectrum and outside it. In the first case ($\cE_{F}^{}\in{\rm
Gap}$), $\sigma_{H}^{}$ is proportional to the Chern number (in the case $T=0$),
which takes on quantized values regardless of the presence or absence of
disorder and/or an external magnetic field. The half-integer ($\rm Ch=\pm 1/2$)
quantization of the anomalous Hall conductance of massive Dirac
electrons\cite{sin_etal_2006,sin_etal_2007,ado_etal_2015,ludw_etal_1994} is 
a~consequence of the fermionic numbers fractionalization\cite{jack_reb_1976,
koen_etal_2014,koen_etal_2016}. It should be emphasized that the bare bubble
part of $\sigma_{H}^{\rm II}$, as well as the contributions to $\sigma_{H}^{}$
due to extrinsic mechanisms, vanish under these conditions. Thus, the survival
of the gap in the electronic spectrum of the Chern insulator is the only
condition for quantizing its anomalous Hall conductance in the presence of
disorder, external magnetic field, and other perturbations.

As an illustration, we calculate in Section \ref{sec:intr_hall} the even in 
a~magnetic field intrinsic anomalous Hall conductance $\sigma_{H}^{\rm int}=
\sigma_{H}^{\rm Ib\,a}+\sigma_{H}^{\rm II\,a}$ of a~two-dimensional disordered 
gas of massive Dirac electrons in SCBA [see Eqs.~(\ref{eq:streda_term_b_0}) and 
(\ref{eq:bare_calc_fin})]. As noted above, the parity of $\sigma_{H}^{\rm a}$
in a magnetic field does not mean the violation of the Onsager relation, since
it requires both normal and anomalous conductances be odd in all parameters
that break the time reversal symmetry of the considered system [see 
Eq.~(\ref{eq:onsager})].

The results of a numerical analysis of the dependencies of
$\sigma_{H}^{\rm II\,a}$ (\ref{eq:streda_term_b_0}), and $\sigma_{H}^{\rm int}=
\sigma_{H}^{\rm Ib\,a}+\sigma_{H}^{\rm II\,a}$ (\ref{eq:streda_term_b_0}), 
(\ref{eq:bare_calc_fin}) on the location of the Fermi level at various values
of the disorder parameter and magnetic field induction are shown in
Fig.~\ref{fig:intr_ahe_chern}. As expected, the quantization of the anomalous 
Hall conductance survives in the presence of sufficiently weak perturbations 
that leave open the gap in the electronic energy spectrum.

Contrary to conventional opinion, the presence of disorder has a~significant 
effect on the $\sigma_{H}^{\rm II}$ behavior beyond the QAHE-plateau (see the
upper left panel in the Fig.~\ref{fig:intr_ahe_chern}). But, as can be seen
from the graphs on the lower left panel of Fig.~\ref{fig:intr_ahe_chern}, the 
corrections to $\sigma_{H}^{\rm II}$ and  $\sigma_{H}^{\rm Ib}$ due to 
scattering of electrons in a~random field of impurities cancel out with a~very 
high precision. In particular, the relative deviation of the sum of these terms 
from its quantized value $\sigma_{0}^{}/2$ for $|\cE|<M$, and from its behavior
in the clean limit
\begin{equation}
\label{eq:int_ahe_clean}
\sigma_{H}^{\rm int}=\sigma_{0}^{}\frac{|\cE M|}{\cE_{}^{2}+M_{}^{2}}
\end{equation}
for $|\cE|>M$ is at least an order of magnitude smaller than the dimensionless 
disorder parameter $\gamma_{0}^{}$.

As already mentioned, the quantization of $\sigma_{H}^{\rm int}$ survives in an
external magnetic field $B$. But, as can be seen from
Fig.~\ref{fig:intr_ahe_chern}, the lower boundary of the Hall plateau shifts 
towards higher energies due to the broadening of the zero Landau level
$\cE_{0,-1}^{}=-M$ as magnetic field $B$ increases. At the same time, the upper
boundary remains almost fixed that leads to a~narrowing of the Hall plateau.

In the region of sufficiently strong magnetic fields, the upper boundary of the 
Hall plateau turns out to be inside the gap of the electron spectrum. In this
case, it represents the threshold at which the $\sigma_{H}^{\rm int}$ as
a~function of the Fermi energy suddenly drops from its quantized value to zero
as in the clean limit. A~similar dependence of the anomalous Hall conductance
on the magnetic field was predicted in Ref.~\onlinecite{boet_etal_2020}. 

The main features of the behavior of the intrinsic anomalous Hall conductance
$\sigma_{H}^{\rm int}$ beyond the plateau are determined by the energy
dependence of the relaxation time $\tau$ (\ref{eq:tau_tr_def}). Indeed, on the
one hand, this parameter plays the role of transport time in the Drude
denominator $1+\omega_{c}^{2}\tau_{}^{2}$ (\ref{eq:bare_calc_fin}), and, on the
other hand, it is related to the Dingle temperature $1/\omega_{c}^{}\tau\propto
k_{B}^{}T_{D}^{}/\hbar\omega_{c}^{}$. Consequently, $\sigma_{H}^{\rm int}$ 
exhibits Shubnikov-de Haas (ShdH) oscillations against the background of 
a~pronounced dip in its energy dependence in vicinity of the plateau if the 
condition of a~strong magnetic field, $\omega_{c}^{}\tau>1$, is satisfied in 
this region. The relaxation time $\tau$ (\ref{eq:tau_tr_def}) is proportional
to $\cE_{F}^{-2}$. As a~result, the amplitudes of ShdH oscillations rapidly
tend to zero, and the conductance asymptotically approaches its value in 
absence of a~magnetic field as the Fermi level increases (See right panel in 
Fig~\ref{fig:intr_ahe_chern}).

Of course, we are talking here about the behavior outside the plateau of only 
intrinsic conductance $\sigma_{H}^{\rm int}$. Calculations of the contributions 
of known external mechanisms to the anomalous Hall conductance of
a two-dimensional gas of massive Dirac fermions in the absence of a magnetic 
field can be found in Refs.~\onlinecite{sin_etal_2006,sin_etal_2007} (side jump
and skew scattering mechanisms) and in Ref.~\onlinecite{ado_etal_2015} 
(coherent skew scattering of electrons by impurity pairs).   

In conclusion, we note that the SCBA used in this article is correct provided 
that $k_{F}^{}l\gg1$, where  $k_{F}^{}$ is the Fermi momentum and $l$ is the 
mean free path. This inequality is violated near the edges of the valence and
conduction bands $|\cE|\simeq M$. In this energy region, SCBA gives only
a~qualitative description of the behavior of the DOSs and Hall conductance.
Besides that, SCBA does not describe the so-called "tails" of the DOS within
the energy gap $|\cE|<M$. Therefore, the question of how the presence of these
"tails" affects the quantization of anomalous Hall conductance remains open.

\begin{acknowledgments}

Author thanks V.\,V. Ustinov, I.\,I.~Lyapilin, and N.\,G. Bebenin for helpful
discussions. This research was carried out within the state assignment of
Ministry of Science and Higher Education of the Russian Federation (theme
"Spin" No. AAAA-A18-118020290104-2).

\end{acknowledgments}

\appendix

\section{\label{app:moyal} Momentum representation in a magnetic field}

An averaged one-electron GF in the spatially uniform disordered
system subject to an external magnetic field has in coordinate representation
$G(\bm r,\bm r')=\cmatrix{\bm r}{G}{\bm r'}$ the following general structure 
\begin{equation}
\label{eq:gf_gen_struct_s}
G(\bm r,\bm r')=e_{}^{i\Phi(\bm r,\bm r_{}')}G(\vrho)=\exp\left[i\frac{e}
{c\hbar}\!\int_{\bm r'}^{\bm r}\!\bm A(\bm x)\cdot d\bm x\right]\! G(\vrho), 
\end{equation}
where $G(\vrho)=G(\bm r-\bm r')$ is a translation invariant and
gauge-independent multiplier, $\Phi(\bm r,\bm r')$ is the gauge-dependent phase, 
and integral in the exponent is calculated along the straight line connecting 
points $\bm r'$ and $\bm r$. For example, in symmetric gauge, this phase is 
equal to $\Phi(\bm r,\bm r')=e\bm B\cdot(\bm r'\times\bm r)/2c\hbar$. Obviously, 
any two-point function invariant in magnetic translations must have such
factorization. For example, the electron self-energy operator in coordinate
representation can be written in the form $\Sigma(\bm r,\bm r')=\exp[i\Phi
(\bm r,\bm r')]\Sigma(\vrho)$. Therefore, given the gauge-independent identity
\begin{equation}
\label{eq:ident_1_s}
e_{}^{-i\Phi(\bm r,\bm r')}\left[\bm p-\frac{e}{c}\bm A(\bm r)\right]
e_{}^{i\Phi(\bm r,\bm r')}=\bm p-\frac{e}{2c}\bm B\times\vrho,
\end{equation}
we obtain equation for the translation invariant multiplier of the one-particle
GF:
\begin{equation}
\label{eq:inv_gf_eq_s}
[\cE-\cH(\vpi)]G(\vrho)-\int e_{}^{i\Theta(\vrhoscr,\vrhoscr')}\Sigma(\vrho
-\vrho')G(\vrho')d\vrho'=\delta(\vrho),
\end{equation} 
where $\cH(\vpi)$ is the Hamiltonian of an electron moving in an external
orthogonal $(\perp\rm OXY)$ magnetic field $(\bm B=\slashnabla\times\bm A)$ ,
$\vpi=-i\hbar\slashnabla_{\!\!\vrhoscr}^{}-e(\bm B\times\vrho)/2c$ is the
operator of gauge-invariant mechanical momentum, $\Theta(\vrho,\vrho')=e\bm B
\cdot(\vrho'\times\vrho)/2c\hbar$ is the phase proportional to the magnetic
flux across the area of triangle built on vectors $\vrho$ and $\vrho'$.

The equation (\ref{eq:inv_gf_eq_s}) is translation invariant that allows us
to passage to the momentum representation with help of Fourier transform
\begin{equation}
\label{eq:fourier_s}
G_{\bm p}^{}=\!\int\! e_{}^{-i\bm p\cdot\vrhoscr/\hbar}G(\vrho)\,d\bm r\,,\quad
G(\vrho)=\!\int\! e_{}^{i\bm p\cdot\vrhoscr/\hbar}G_{\bm p}^{}\frac{d\bm p}
{(2\pi\hbar)_{}^{2}}.
\end{equation}
As a result, we obtain the equation
\begin{equation}
\label{eq:gf_eq_p_s}
\left[\cE-\cH\left(\bm p-i\frac{e\hbar}{2c}\bm B\times\slashnabla_{\!\!\bm p}^{}
\right)\right]G_{\bm p}^{}-\Sigma_{\bm p}^{}\star G_{\bm p}^{}=1,
\end{equation}
where symbol $F\star G$ denotes the Moyal
$\star$-product\cite{moyal_1949,zach_etal_2005} of the functions dependent on
the momentum $\bm p$
\begin{equation}
\label{eq:star_def_s_1}
F_{\bm p}^{}\star G_{\bm p}^{}=\frac{l_{B}^{4}}{\pi_{}^{2}}\!\iint\!
F_{\bm p+\hbar\bm k}^{}e_{}^{-i2l_{B}^{2}\bm b\cdot(\bm k\times\bm k')}
G_{\hbar\bm k'+\bm p}^{}\,d\bm k\,d\bm k',
\end{equation}
where $l_{B}^{}=\sqrt{c\hbar/|e|B}$ is the magnetic length and $\bm b=\bm B/B$
is the unit vector parallel to the magnetic field $\bm B$. In the general case,
the $\star$-product (\ref{eq:star_def_s_1}) is non-commutative, but it
satisfies the requirements of associativity and distributivity. An useful 
representation of the $\star$-product can be obtained by rewriting 
(\ref{eq:star_def_s_1}) with help of the argument shift operator
$F_{\bm p+\hbar\bm k}^{}=\exp(\hbar\bm k\cdot\slashnabla_{\!\!\bm p}^{})
F_{\bm p}^{}$. After calculating the resulting Gaussian integral, we obtain the
definition of the $\star$-product (\ref{eq:star_product}), which is equivalent
to (\ref{eq:star_def_s_1}). Since $\star$-product (\ref{eq:star_product})
includes an exponential function dependent on the gradient operator, it can be 
represented through shift of the argument in one of its multipliers, for example,
\begin{equation}
\label{eq:star_def_s_3}
F_{\bm p}^{}\star G_{\bm p}^{}=F\left(\bm p-i\frac{e\hbar}{2c}\bm B\times
\stackrel{\rightarrow}\slashnabla_{\!\!\bm p}^{}\right)G_{\bm p}^{}.
\end{equation}

Given the last definition of the $\star$-product, we can rewrite the equation
(\ref{eq:gf_eq_p_s}) in the form used in main text [See
Eq.~(\ref{eq:dyson_p_repres})]. It looks almost like an equation for the
averaged one-particle GF in absence of a magnetic field. The only
difference is that (\ref{eq:dyson_p_repres}) contains the $\star$-product
instead of usual multiplication. The momentum determined by the Fourier
transform (\ref{eq:fourier_s}) has the meaning of gauge-invariant mechanical
momentum, which is proportional to the electron velocity, i.e., $\bm p=m\bm v$.
Indeed the turning on a magnetic field results in this representation to
replacing conventional products by the $\star$-products of momentum dependent
functions $F_{\bm p}^{}G_{\bm p}^{}\mapsto F_{\bm p}^{}\star G_{\bm p}^{}$. In
particularly, $p_{\alpha}^{}p_{\beta}^{}\,\mapsto\,p_{\alpha}^{}\star
p_{\beta}^{}=p_{\alpha}^{}p_{\beta}^{}-i\epsilon_{\alpha\beta}^{}\hbar_{}^{2}/2
l_{B}^{2}$ ($\alpha(\beta)=x,y$). It follows that the $\star$-commutator of the
Cartesian components of the momentum is equal to $\big[p_{x}^{}\star p_{y}^{}
\big]=p_{x}^{}\star p_{y}^{}-p_{y}^{}\star p_{x}^{}=-i\hbar_{}^{2}/l_{B}^{2}$.  
Thus, the algebra of the momentum dependent $c$-numerical functions with
respect to $\star$-multiplication is equivalent to the algebra of the mechanical
momentum operators of an electron in a~magnetic field.

The transition to the Fourier transform in a magnetic field is useful, since
$G_{\bm p}^{}$ possesses some properties of the one-electron GF in the momentum
representation. In particularly, the trace over spatial degrees of freedom is
expressed through integral with respect to momentum
\begin{equation}
\label{eq:trace_formula_s}
\Sp G=\int\!G_{\bm p}^{}\frac{d\bm p}{(2\pi\hbar)_{}^{2}}\,.
\end{equation} 
A similar representation also holds for the trace of the two-particle
correlation function $\Sp\big\langle V_{\alpha}^{}\sG_{}^{R}V_{\beta}^{}
\sG_{}^{A}\big\rangle$ through which the components of the electrical 
conductivity tensor are expressed.


\begin{thebibliography}{99}
\bibitem{footnote_1} Here and below we drop superscripts $R(A)$ if this does
not lead to misunderstandings; sometimes, for brevity of notation, we shall not
write explicitly the energy arguments of GFs and electron self-energies.
\end{thebibliography}
\end{document}